\documentclass{emulateapj}
\usepackage{apjfonts}
\newcommand{\etal}{et~al.}
\newcommand{\PVdblt}{{\rm P}\kern 0.1em{\sc v}~$\lambda\lambda 1117, 1128$}
\newcommand{\CaIIdblt}{{\rm Ca}\kern 0.1em{\sc ii}~$\lambda\lambda 3934, 3969$}
\newcommand{\AlIIIdblt}{{\rm Al}\kern 0.1em{\sc iv}~$\lambda\lambda 1855, 1863$}
\newcommand{\CIVdblt}{{\rm C}\kern 0.1em{\sc iv}~$\lambda\lambda 1548, 1550$}
\newcommand{\MgIIdblt}{{\rm Mg}\kern 0.1em{\sc ii}~$\lambda\lambda 2796, 2803$}
\newcommand{\NVdblt}{{\rm N}\kern 0.1em{\sc v}~$\lambda\lambda 1238, 1242$}  
\newcommand{\SVIdblt}{{\rm S}\kern 0.1em{\sc vi}~$\lambda\lambda 933, 944$} 
\newcommand{\OVIdblt}{{\rm O}\kern 0.1em{\sc vi}~$\lambda\lambda 1031, 1037$} 
\newcommand{\SiIIdblt}{{\rm Si}\kern 0.1em{\sc ii}~$\lambda\lambda 1190, 1193$} 
\newcommand{\SiIVdblt}{{\rm Si}\kern 0.1em{\sc iv}~$\lambda\lambda 1393, 1402$} 
\newcommand{\PV}{\hbox{{\rm P}\kern 0.1em{\sc v}}}
\newcommand{\AlI}{\hbox{{\rm Al}\kern 0.1em{\sc i}}}
\newcommand{\AlII}{\hbox{{\rm Al}\kern 0.1em{\sc ii}}}
\newcommand{\AlIII}{{\hbox{\rm Al}\kern 0.1em{\sc iii}}}
\newcommand{\CaII}{\hbox{{\rm Ca}\kern 0.1em{\sc ii}}}
\newcommand{\CII}{\hbox{{\rm C}\kern 0.1em{\sc ii}}}
\newcommand{\CIIe}{\hbox{{\rm C$^{\ast}$}\kern 0.1em{\sc ii}}}
\newcommand{\CIII}{\hbox{{\rm C}\kern 0.1em{\sc iii}}}
\newcommand{\CIV}{\hbox{{\rm C}\kern 0.1em{\sc iv}}}
\newcommand{\CV}{\hbox{{\rm C}\kern 0.1em{\sc v}}}
\newcommand{\HI}{\hbox{{\rm H}\kern 0.1em{\sc i}}}
\newcommand{\HII}{\hbox{{\rm H}\kern 0.1em{\sc ii}}}
\newcommand{\Lya}{\hbox{{\rm Ly}\kern 0.1em$\alpha$}}
\newcommand{\Lyb}{\hbox{{\rm Ly}\kern 0.1em$\beta$}}
\newcommand{\Lyg}{\hbox{{\rm Ly}\kern 0.1em$\gamma$}}
\newcommand{\Lyd}{\hbox{{\rm Ly}\kern 0.1em$\delta$}}
\newcommand{\Lye}{\hbox{{\rm Ly}\kern 0.1em$\epsilon$}}
\newcommand{\Lyphi}{\hbox{{\rm Ly}\kern 0.1em$\phi$}}
\newcommand{\Lyfive}{\hbox{{\rm Ly}\kern 0.1em$5$}}
\newcommand{\Lysix}{\hbox{{\rm Ly}\kern 0.1em$6$}}
\newcommand{\Lyseven}{\hbox{{\rm Ly}\kern 0.1em$7$}}
\newcommand{\Lyeight}{\hbox{{\rm Ly}\kern 0.1em$8$}}
\newcommand{\Lynine}{\hbox{{\rm Ly}\kern 0.1em$9$}}
\newcommand{\Lyten}{\hbox{{\rm Ly}\kern 0.1em$10$}}
\newcommand{\Lyeleven}{\hbox{{\rm Ly}\kern 0.1em$11$}}
\newcommand{\HeI}{\hbox{{\rm He}\kern 0.1em{\sc i}}}
\newcommand{\HeII}{\hbox{{\rm He}\kern 0.1em{\sc ii}}}
\newcommand{\FeI}{\hbox{{\rm Fe}\kern 0.1em{\sc i}}}
\newcommand{\FeII}{\hbox{{\rm Fe}\kern 0.1em{\sc ii}}}
\newcommand{\FeIII}{\hbox{{\rm Fe}\kern 0.1em{\sc iii}}}
\newcommand{\MnII}{\hbox{{\rm Mn}\kern 0.1em{\sc ii}}}
\newcommand{\MgI}{\hbox{{\rm Mg}\kern 0.1em{\sc i}}}
\newcommand{\MgII}{\hbox{{\rm Mg}\kern 0.1em{\sc ii}}}
\newcommand{\MgIII}{\hbox{{\rm Mg}\kern 0.1em{\sc iii}}}
\newcommand{\NI}{\hbox{{\rm N}\kern 0.1em{\sc i}}}
\newcommand{\NII}{\hbox{{\rm N}\kern 0.1em{\sc ii}}}
\newcommand{\NIII}{\hbox{{\rm N}\kern 0.1em{\sc iii}}}
\newcommand{\NV}{\hbox{{\rm N}\kern 0.1em{\sc v}}}
\newcommand{\OVI}{\hbox{{\rm O}\kern 0.1em{\sc vi}}}
\newcommand{\OI}{\hbox{{\rm O}\kern 0.1em{\sc i}}}
\newcommand{\OII}{\hbox{[{\rm O}\kern 0.1em{\sc ii}]}}
\newcommand{\OIII}{\hbox{[{\rm O}\kern 0.1em{\sc iii}]}}
\newcommand{\OIV}{\hbox{{\rm O}\kern 0.1em{\sc iv}]}}
\newcommand{\SI}{{\rm S}\kern 0.1em{\sc i}}
\newcommand{\SIV}{{\rm S}\kern 0.1em{\sc iv}}
\newcommand{\SVI}{{\rm S}\kern 0.1em{\sc vi}}
\newcommand{\SiI}{\hbox{{\rm Si}\kern 0.1em{\sc i}}}
\newcommand{\SiII}{\hbox{{\rm Si}\kern 0.1em{\sc ii}}}
\newcommand{\SiIII}{\hbox{{\rm Si}\kern 0.1em{\sc iii}}}
\newcommand{\SiIV}{\hbox{{\rm Si}\kern 0.1em{\sc iv}}}
\newcommand{\SII}{\hbox{{\rm S}\kern 0.1em{\sc ii}}}
\newcommand{\SIII}{\hbox{{\rm S}\kern 0.1em{\sc iii}}}
\newcommand{\NaI}{\hbox{{\rm Na}\kern 0.1em{\sc i}}}
\newcommand{\TiII}{\hbox{{\rm Ti}\kern 0.1em{\sc ii}}}
\newcommand{\kms}{\hbox{km~s$^{-1}$}}
\newcommand{\cmsq}{\hbox{cm$^{-2}$}}

\slugcomment{ApJ Lett.: Accepted 11 Nov 2011}
\shorttitle{\sc Outflow and Infall {\HI} Content}
\shortauthors{\sc Kacprzak \& Churchill}

\begin{document}


\title{The {\HI} Mass Density in Galactic Halos, Winds, and Cold Accretion \\
as Traced by {\MgII} Absorption}


\author{\sc
Glenn G. Kacprzak\altaffilmark{1,2}
and
Christopher W. Churchill\altaffilmark{3,4}
}
                                                                                
\altaffiltext{1}{Swinburne University of Technology, Victoria 3122,
Australia {\tt gkacprzak@astro.swin.edu.au}}

\altaffiltext{2}{Australian Research Council Super Science Fellow}

\altaffiltext{3}{New Mexico State University, Las Cruces, NM 88003
{\tt cwc@nmsu.edu}}

\altaffiltext{4}{Visiting Professor, Swinburne University of
Technology}

\begin{abstract}

It is well established that {\MgII} absorption lines detected in
background quasar spectra arise from gas structures associated with
foreground galaxies.  The degree to which galaxy evolution is driven
by the gas cycling through halos is highly uncertain because their gas
mass density is poorly constrained.  Fitting the {\MgII} equivalent
width ($W$) distribution with a Schechter function and applying the
$N({\HI})$--$W$ correlation of M{\'e}nard \& Chelouche, we computed
$\Omega(\HI)_{\hbox{\tiny \MgII}} \equiv \Omega ({\HI})_{\rm halo} =
1.41^{+0.75}_{-0.44}\times 10^{-4}$ for $0.4 \leq z \leq 1.4$.  We
exclude DLAs from our calculations so that $\Omega$(\HI)$_{\rm halo}$
comprises accreting and/or outflowing halo gas not locked
up in cold neutral clouds.  We deduce the cosmic {\HI} gas mass
density fraction in galactic halos traced by {\MgII} absorption is
$\Omega ({\HI})_{\rm halo}/\Omega ({\HI}) _{\hbox{\tiny DLA}} \simeq
15$\% and $\Omega ({\HI})_{\rm halo}/\Omega_b \simeq 0.3$\%.  Citing
several lines of evidence, we propose infall/accretion material is
sampled by small $W$ whereas outflow/winds are sampled by large $W$,
and find $\Omega(\HI)_{\rm infall}$ is consistent with
$\Omega(\HI)_{\rm outflow}$ for bifurcation at
$W=1.23^{+0.15}_{-0.28}$~{\AA}; cold accretion would then comprise no
more than $\sim 7$\% of of the total {\HI} mass density.  We discuss
evidence that (1) the total {\HI} mass cycling through halos remains
fairly constant with cosmic time and that the accretion of {\HI} gas
sustains galaxy winds, and (2) evolution in the cosmic star formation
rate depends primarily on the {\it rate\/} at which cool {\HI} gas
cycles through halos.

\end{abstract}




\section{Introduction}
\label{sec:intro}

Our knowledge of galaxy evolution relies heavily on both observations
and simulations that focus on the mechanisms by which galaxies
acquire, chemically enrich, recycle, and expel their gaseous
component.  However, the mean {\it quantity\/} of halo gas engaged in
any given process remains poorly constrained. The use of quasar
absorption lines provides a unique tool to directly observe these
ongoing processes and allows for sensitive measures of the quantity of
gas within galaxy halos along with their cross-sections, kinematics,
metallicities, densities, and temperatures.

The dense neutral hydrogen that is mostly confined within galaxies
commonly exhibits damped {\Lya} absorption [$N(\HI) \geq
2\times10^{20}$ cm$^{-2}$ $\equiv \hbox{DLA}$].  DLAs are
fundamentally different from other classes of absorption systems. For
example, the Lyman limit systems [$10^{17.3} < N({\HI}) < 10^{20.3}$
cm$^{-2}$ $\equiv$ LLS] have varying degrees of hydrogen ionization
\citep[cf.,][]{prochaska99} and extend out to $\sim 100$~kpc around
galaxies\footnote{LLS have $d\aleph/dz$ consistent with that of {\MgII}
absorbers \citep{stengler-kp95,nestor05}, which have $R\simeq 100$~kpc
\citep{kacprzak08,chen10a}.}, whereas DLAs are believed to probe the
cool, dense precursors of star forming molecular clouds and can
account for up to $\sim 50$\% of the galactic baryonic content
\citep[cf.,][]{wolfe+05}.

In an effort to determine the quantity of dense gas within galaxies,
\citet{rao06} measured the cosmological neutral gas mass density
traced by DLAs to be $\Omega(\HI)_{\hbox{\tiny DLA}}= (9.6 \pm 4.5)
\times 10^{-4}$ at $<z>=0.92$.  It has also been argued that
$\Omega(\HI)_{\hbox{\tiny DLA}}$ remains roughly constant for $z \sim
0.2-5$ \citep{prochaska04,peroux05,rao06,lah08,songaila10,meiring11}
and then decreases by a factor of $\sim 2-3$ by $z=0$
\citep{zwaan05a,martin10}.  There is also some evidence showing that
$\Omega(\HI)_{\hbox{\tiny DLA}}$ might decrease below $z=2.2$
\citep{noterdaeme09}.

The global star formation rate (SFR) history of the universe has
evolved dramatically \citep[e.g.,][]{bouwens11} and if the redshift
constancy of $\Omega (\HI)_{\hbox{\tiny DLA}}$ holds, then this
implies that DLA gas does not directly track the formation of stars.
This might imply that the global SFR is predominantly governed by
mechanisms linked to galactic halos, such as gas accretion from the
intergalactic medium and/or recycling of gas within the galaxy halos.
An estimate of $\Omega(\HI)_{\rm halo}$ ({\it excluding} {\HI} from
DLAs), being the sum of an accreting/infall component
$[\Omega(\HI)_{\rm infall}]$ and a wind/outflow component
$[\Omega(\HI)_{\rm outflow}]$, could place constraints on the relative
importance and roles with which these processes drive star formation
in galaxies.

The {\MgIIdblt} absorption doublet, which probes low-ionization gas
with $10^{16} \leq N(\HI) \leq 10^{22}$~{\cmsq}, is commonly used to
study the gaseous components of galaxies \citep[see][for a
review]{cwc-china}. The {\MgII} absorption is observed out to
projected galactic radii of $\sim 100$ kpc \citep{kacprzak08,chen10a}.

For {\MgII} absorption systems with rest-frame equivalent widths of $W
\geq 1$~{\AA}, galaxy color and star formation rate correlates
strongly with $W$ \citep{zibetti07,noterdaeme10,nestor11}, a result
highly suggestive that galactic outflows are responsible for ejecting
substantial amounts of gas to large galactocentric radii.  The outflow
scenario is also supported by the result of \citet{bouche06}, who
found an anti-correlation between $W$ and the host halo mass, by
cross-correlating {\MgII} absorbers with luminous red galaxies from
SDSS, and claim that this provides evidence that absorbers are not
virialized in gaseous halos of the galaxies. They suggest that the
strong absorbers are statistically more likely to trace super-winds.

Indeed, direct evidence for {\MgII} absorbing winds is seen in spectra
of star forming galaxies, which exhibit strong outflows blueshifted
$300-1000$~{\kms} relative to the galaxy
\citep{tremonti07,weiner09,martin09,rubin10b}.  These galaxies almost
exclusively exhibit $W \geq 1$~{\AA} absorption. \citet{chelouche10}
demonstrated that models of outflowing wind-driven gas reproduce the
{\MgII} velocity widths of $W\geq 1$~{\AA} systems observed with high
resolution. The extent of these {\MgII} absorbing winds are not well
constrained; however, \citet{bordoloi11} suggest they may reach out to
$\sim 50$~kpc.

For samples dominated by $W < 1$~{\AA}, neither \citet{chen10a} nor
\citet{kacprzak11b} found a $W$--galaxy color correlation, contrary to
the \citet{zibetti07} result for $W > 1$~{\AA}.  Furthermore,
\citet{chen10b} found that {\MgII} ``halo size'' increases with
increasing galaxy stellar mass and weakly with specific star formation
rate, suggesting a scenario in which infalling {\MgII} absorbing gas
structures (selected by $W<1$~{\AA} absorption) fuel star formation.

The SPH simulations of \citet{stewart11a} reveal that gas-rich mergers
and cold-flow streams produce a circumgalactic co-rotating,
low-ionization gas component that is predominately infalling towards
the galaxy.  In absorption, these structures are expected to exhibit
$\sim 100$~{\kms} velocity offsets relative to the host galaxy,
consistent with the observations of \citet{steidel02},
\citet{kacprzak10a}, and \citet{kacprzak11a}.  This spatial/kinematic
configuration yields a correlation between galaxy inclination and $W$,
which has been previously observed by \citet{kacprzak11b}.

The above body of evidence suggest that weaker {\MgII} absorption
selects gas accretion from infalling cold streams or cooled gas
returning from earlier processing within the galaxy, whereas stronger
absorption ($W \geq 1$~{\AA}) selects outflows.  It would be useful to
constrain the relative neutral gas mass density for both processes,
i.e., $\Omega(\HI)_{\rm halo} = \Omega(\HI)_{\rm infall} +
\Omega(\HI)_{\rm outflow}$, to gain insight into how much galactic gas
is cycled through either mechanism at a given time and how this
compares to $\Omega (\HI)_{\hbox{\tiny DLA}}$.

In this {\it Letter}, we compute the {\HI} mass density within galaxy
halos traced by {\MgII} absorption, $\Omega(\HI)_{\rm halo}\equiv
\Omega ({\HI}) _ {\hbox{\tiny \MgII}}$. We apply the $N({\HI})$--$W$
relation of \citet{menard09b} to obtain the {\HI} column density
distribution function directly from the {\MgII} equivalent width
distribution function.  We exclude DLAs from our calculations so that
$\Omega$(\HI)$_{\rm halo}$ budgets gas likely to be accreting and/or
outflowing from galaxies but not locked up in cold neutral clouds.  We
apply a 1~{\AA} bifurcation to compute $\Omega (\HI)_{\rm infall}$ and
$\Omega(\HI)_{\rm outflow}$, and determine the $W$ at which
$\Omega(\HI)_{\rm infall} = \Omega(\HI)_{\rm outflow}$.  Throughout we
adopt a $h=0.70$, $\Omega_{\rm M}=0.3$, $\Omega_{\Lambda}=0.7$
cosmology.

\section{Computing $\Omega({\HI})$ from {\MgII}}
\label{sec:omg}

The cosmological neutral hydrogen gas mass density is computed from
\begin{equation}
\Omega ({\HI}) = \frac{H_0}{c} \frac{\mu m_{\hbox{\tiny H}}}{\rho_c} 
\left< N(\hbox{\HI}) \right> \frac{d\aleph}{dz}\frac{E(z)}{(1+z)^2} , 
\label{eq:omg}
\end{equation}
given that,
\begin{equation}
 \left< N(\hbox{\HI}) \right> \frac{d\aleph}{dz}=\int_0^\infty \!\!\!\! N n(N) \, dN ,
\label{eq:nHI}
\end{equation}
where $n(N)$ is the {\HI} column density distribution function, $\mu =
1.3$ is the mean molecular weight\footnote{$\mu = 1.3$ applies for a
  fully neutral gas and is the value used for DLA studies.  For
  partially ionized gas, $\mu$ is slightly smaller, but still of order
  unity.}, $\rho_c$ is the critical density of the universe, $H_0$ is
the Hubble constant, $m_{\hbox{\tiny H}}$ is the mass of hydrogen,
$\left< N({\HI}) \right>$ is the mean {\HI} column density, $d\aleph/dz$ is
the number of systems per unit redshift, and $E(z)= H(z)/H_0 =
[\Omega_{\hbox{\tiny M}}(1+z)^3+\Omega_{\Lambda}]^{1/2}$.

%%%%%%%%%%%%%%%%%%%%%%%%%%%%%%%%%%%%%%%%%%%%%%%%%%%%%%%%%%%%%%%%%%
\begin{figure}[th]
%\vglue -0.25in
\includegraphics[angle=0,scale=0.59]{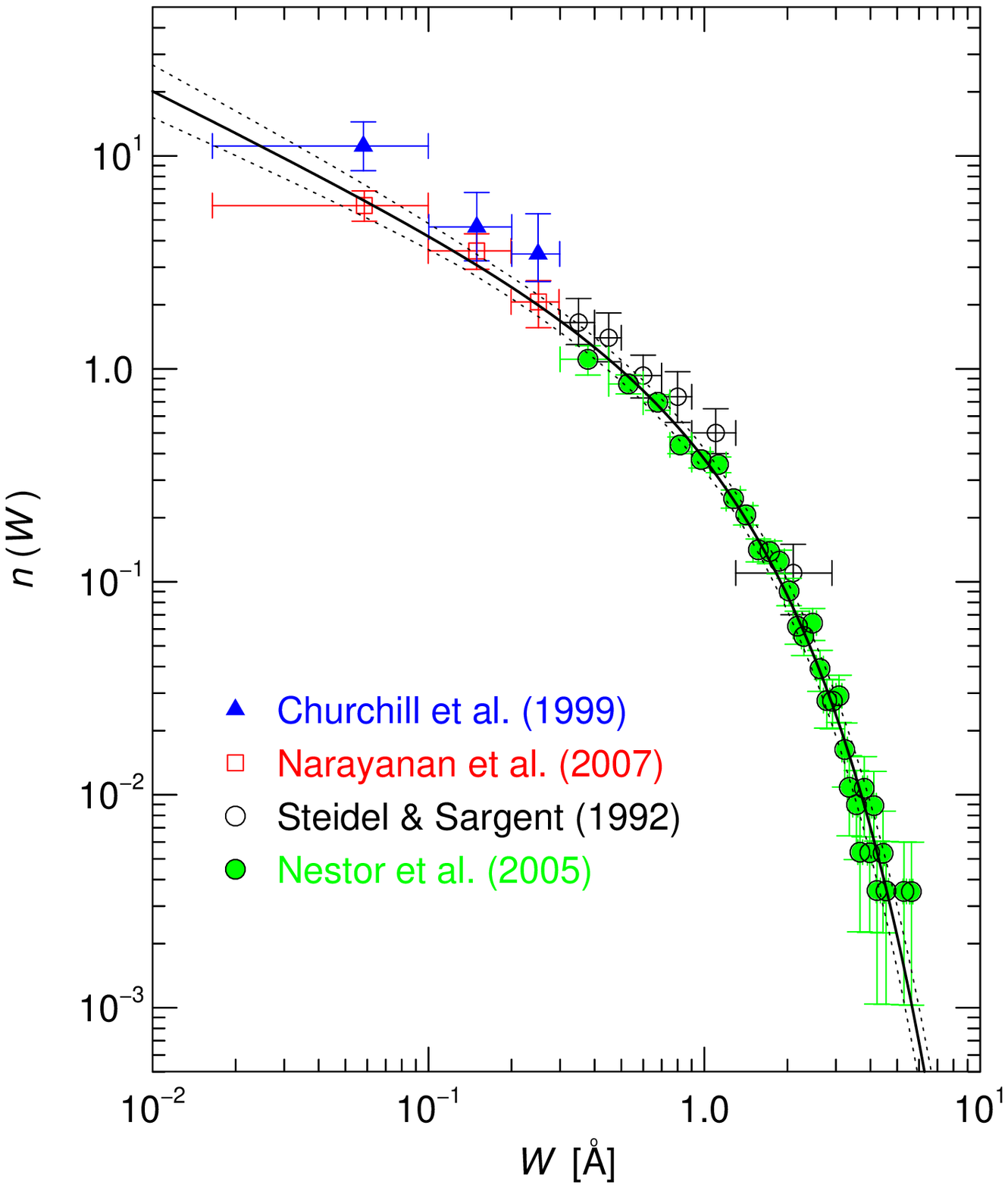}
\caption[angle=0]{The distribution of {\MgII} rest-frame equivalent
  widths, $n(W)$, using the data from \citet{weakI} [blue triangles],
  \citet{narayanan07} [red open squares], \citet{steidel92} [black
  open circles] and \citet{nestor05} [green circles]. We fitted the
  data using a Schechter function (solid curve) with parameters
  $C_*=1.08\pm 0.12$~{\AA}$^{-1}$, $\alpha=-0.642\pm0.062$ and $W_* =
  0.970\pm 0.056$~{\AA}. The dashed curves represent the maximum
  $1~\sigma$ error in the fit for the parameter uncertainties.}
\label{fig:nW}
%\vglue -0.05in
\end{figure}
%%%%%%%%%%%%%%%%%%%%%%%%%%%%%%%%%%%%%%%%%%%%%%%%%%%%%%%%%%%%%%%%%%

Computing $\Omega ({\HI})$ traced by {\MgII} absorption requires
$n(N)$, which is not directly observationally known for {\MgII}
absorbers.  However, \citet{menard09b} determined the geometric mean
column density as a function of $W$ to be $N({\HI})= A W^{\beta}$,
where $A=(3.06 \pm 0.55) \times 10^{19}$ cm$^{-2}$~{\AA}$^{-\beta}$
and $\beta=1.73\pm 0.26$ for $0.5 \leq W \leq 3$~{\AA} and $0.5 \leq z
\leq 1.4$.  Employing this column density relation, we can obtain
$n(N)dN$ from the equivalent width distribution, $n(W)dW$, where
$n(W)$ is the distribution of {\MgII} rest-frame equivalent widths.

In Figure~\ref{fig:nW}, we plot $n(W)$ versus $W$ using the data of
\citet{steidel92}, \citet{weakI}, \citet{nestor05}, and
\citet{narayanan07}.  Since the work of \citet{steidel92}, it has been
common to fit $n(W)$ as either a power law or an exponential
function. For $W\geq0.3$~{\AA}, \citet{steidel92} could not
distinguish which function was preferred, whereas \citet{nestor05}
showed that the distribution was clearly exponential. For
$W\leq0.3$~{\AA}, \citet{weakI} showed that $n(W)$ is a power law that
is consistent with the \citet{steidel92} data for $W\leq1.0$~{\AA},
which was confirmed by \citet{narayanan07}.

We used $\chi$-squared minimization to fit a \citet{schechter76}
function to the binned data,
\begin{equation}
n(W)dW= C_* \left(\frac{W}{W_*}\right)^\alpha \!\!
\exp \left\{ -\frac{W}{W_*} \right\} d\!\left( \frac{W}{W_*} \right) ,
\label{eq:nW}
\end{equation}
where $C_*$ is the normalization such that the unweighted integral is
equal to $d\aleph/dz$ (to satisfy Eq.~\ref{eq:nHI}).  In
Figure~\ref{fig:nW}, we show our fit with parameters $C_*=1.08\pm
0.12$~{\AA}$^{-1}$, $\alpha=-0.642\pm0.062$ and $W_* = 0.970\pm
0.056$~{\AA}. The {\it overall\/} equivalent width distribution at
$\left< z \right> = 1.0$ is much better fit with a Schechter function
than the power law only \citep[see Fig.~6 of][]{narayanan07} or with
the exponential only \citep[see Fig.~20 of][]{nestor05}.
Interestingly, the characteristic equivalent width, $W_*$, marking the
transition from a power law to an exponential distribution, is
consistent with 1~{\AA}. Based upon the results discussed in
\S~$\!\!$\ref{sec:intro}, we speculate the break in the $n(W)$
distribution at $W\simeq 1$~{\AA} is physical, that infall/accretion
structures yield a power law distribution whereas outflowing/wind
structures yield an exponential distribution.

We note there are subtle differences between the data sets shown in
Figure~\ref{fig:nW}.  The $W<0.3$~{\AA} data of \citet{weakI} [30
systems] and \citet{narayanan07} [112 systems] represent $0.4 \leq z
\leq 1.4$ and have mean redshift of $\left<z\right>=0.9$.  The
$W\geq0.3$~{\AA} data of \citet{steidel92} [103 systems] represent
$0.2\leq z \leq 2.2$ with $\left<z\right>=1.1$, and the data of
\citet{nestor05} [1331 systems] represent $0.4\leq z \leq 2.3$, also
with $\left<z\right>=1.1$.  Thus, the redshift ranges of the low and
high equivalent width samples are not identical, even though the
$\left<z\right>$ are fairly consistent.  Both \citet{steidel92} and
\citet{nestor05} demonstrated redshift evolution of $n(W)$ in that
there is more power at large $W$ at higher redshifts.  This evolution
is expected to result in a slightly larger $W_*$ than if we confined
their samples to $z\leq 1.4$ (the upper limit of the $W<0.3$~{\AA}
data).  Given the mean redshifts of both of the $W\geq 0.3$ ~{\AA}
samples, $\left<z\right>=1.1$, are similar to those of the
$W<0.3$~{\AA} samples, $\left<z\right>=0.9$, we expect that this
slightly increased power in $n(W)$ has only a minor influence on our
final $\Omega ({\HI})_{\rm halo}$ result, as compared to, for example,
the uncertainty in the fitted values of $A$ and $\beta$ in the column
density relation of \citet{menard09b}. We also note (see
Figure~\ref{fig:nW}) a slight vertical offset between the two
$W<0.3$~{\AA} data sets and between the two $W\geq0.3$~{\AA} data
sets.  In both cases the larger surveys yield slightly smaller
$d\aleph/dz$ and the source of the offsets is still unknown.

We now substitute $N({\HI})= A W^{\beta}$ into
Eqs.~\ref{eq:omg}--\ref{eq:nW} and derive,
\begin{equation}
\Omega(\hbox{\HI})= 
\frac{H_0}{c}
\frac{\mu m_{\hbox{\tiny H}}}{\rho_c}
\frac{E(z)}{(1+z)^2} 
\,  C_*A \, W_*^{\beta +1} \, \Gamma (a,w) , 
\label{eq:OHI}
\end{equation}
where $\Gamma(a,w)$ is the incomplete gamma function, and where $a =
\alpha+\beta+1$.  The value of $w$ allows integration over specific
$W$ intervals: $w=W_{max}/W_*$ for integration $0 \rightarrow w$, and
$w=0$ or $W_{min}/W_*$ for integration $w \rightarrow \infty$, where
$W_{min}$ and $W_{max}$ are selected cutoffs.

To compute Eq.~\ref{eq:OHI}, we are required to extrapolate the column
density relation of \citet{menard09b} to $W$ values both lower and
higher than the domain of their fit. Though there is scatter in the
column density relation, it correctly predicts that $W < 0.3$~{\AA}
absorption systems are sub-LLS with $\log N({\HI}) < 17$, consistent
with the observations of \citet{archiveI}.

Since we desire to compute the {\HI} mass density in galaxy halos, we
correct for the quantity of {\HI} found in disks.  The majority of
DLAs lie within the optical disk and exhibit covering fractions of
virtually 100\% \citep[e.g.,][]{zwaan05b}, thus correcting for the
{\HI} DLA absorption provides a good approximation for the disk
contribution of {\HI}.  Though some DLAs may arise in halos, this
fraction is estimated to be less than 1--5\% \citep{fumagalli11}.

\citet{rao06} showed that the probability of a {\MgII} selected DLA
system is $P(W)=0$ for $W< 0.6$~{\AA} and then increases with
increasing $W$ for $W \geq 0.6$~{\AA}.  Using a maximum-likelihood fit
to the binned data in their Fig.~4, we estimate this increase as a
linear function $P(W) = 0.23 W - 0.057$ for $0.6 \leq W \leq
4.5$~{\AA} with $P(W)=1$ for $W > 4.5$~{\AA}.  In order to correct for
DLA contamination in our calculation, we weight Eq.~\ref{eq:OHI} by
$P(W)-1$.  The upper limit of $W=4.5$~{\AA} where DLA contamination is
100\%, is consistent with the \citet{menard09b} column density
relation, which predicts $\log N({\HI}) > 20.3$ for $W\geq4.5$~{\AA}.

\section{Results and Discussion}

In Table~\ref{tab:OmegaHI}, we present $\Omega ({\HI})$ for selected
$W$ ranges.  The quoted uncertainties are the $1~\sigma$ confidence
levels based upon the uncertainties in the fitted parameters $C_*$,
$\alpha$, $W_*$, $A$, and $\beta$.  We deduce the {\HI} mass density
traced by {\MgII} absorption, which is interpreted as the diffuse
{\HI} contained within galaxy halos, is $\Omega ({\HI})_{\rm halo} =
1.41^{+0.75}_{-0.44}\times 10^{-4}$. This value is $\sim 15\%$ of
$\Omega ({\HI})_{\hbox{\tiny DLA}}$, indicative that a considerable
fraction of {\HI} is contained in galaxy halos relative to the {\HI}
in DLAs; it contributes 0.3\% to the total baryonic budget
\citep[$\Omega_b = 0.045$,][]{wmap-7yr} at $\left<z\right>=1.0$.

We find that $W\leq 0.3$~{\AA} {\MgII} absorption \citep[often called
``weak'' systems, e.g.,][]{weakI} selects a small fraction of the
{\HI} mass density, $\Omega ({\HI})^{(<0.3)}_{\hbox{\tiny \MgII}} =
7.71^{+8.71}_{-3.88}\times 10^{-6}$.  From this quantity, it is
difficult to ascertain what fraction of these systems could be
selecting {\Lya} forest structures versus galactic halo structures
because estimates of $\Omega_{\hbox{\tiny \Lya}}$ in the appropriate
$N({\HI})$ range ($10^{15.5}$--$10^{16.5}$~{\cmsq}) are highly
uncertain and are quoted in units of {\it total\/} gas mass density
\cite[neutral + ionized, cf.,][]{penton04}.  The simulations of
\citet{dave10} indicate that $\Omega({\HI}) \simeq 10^{-7}$ for
$N({\HI}) \leq 10^{15}$ (outside halos) at $z\sim 0.1$, where the
{\HI} fraction in the {\Lya} forest is at its highest.

\begin{deluxetable}{lcccl}
\tabletypesize{\scriptsize} 
\tablecaption{ $\Omega ({\HI})$ traced by {\MgII} 
at $0.4 \leq z \leq 1.4$\label{tab:OmegaHI}}
\tablecolumns{3} 
\tablewidth{0pt}
\tablehead{
\colhead{ }                                                &
\colhead{$\Omega(\hbox{\HI})$}                             &
\colhead{$\Omega({\HI})/\Omega({\HI})_{\hbox{\tiny DLA}}$} &
\colhead{$W$ range [\AA]} 
}
\startdata
$\Omega({\HI})_{\rm halo}$                  & $1.41^{+0.75}_{-0.44}\times 10^{-4}$ & 0.147 & $0.0-\infty$\tablenotemark{ a}  \\[1.5ex]
$\Omega({\HI})_{\rm infall}$                & $5.56^{+2.60}_{-1.54}\times 10^{-5}$ & 0.058 & $0.0-1.0$    \\[1.5ex]
$\Omega({\HI})_{\rm outflow}$               & $8.57^{+0.86}_{-1.10}\times 10^{-5}$ & 0.089 & $1.0-\infty$\tablenotemark{ a} \\[1.5ex]
$\Omega({\HI})^{(<0.3)}_{\hbox{\tiny \MgII}}$  & $7.71^{+8.71}_{-3.88}\times 10^{-6}$ & 0.008 & $0.0-0.3$    \\[-1.5ex]
\enddata
\tablenotetext{a}{For $W > 4.5$~{\AA}, DLA contamination is 100\% and
the contribution to $\Omega ({\HI}) _{\rm halo}$ vanishes. See text for
further discussions.}
\end{deluxetable}

We previously described the theoretical and observational evidence
supporting the idea that weaker {\MgII} systems trace infall/accretion
and stronger systems trace outflow/winds and that $W_* \simeq 1$~{\AA}
marks the transition between the two regimes. Applying a $1$~{\AA}
bifurcation to $\Omega({\HI})_{\rm halo}$, we find $\Omega
({\HI})_{\rm infall} = 5.56^{+2.60}_{-1.54}\times 10^{-5}$ and $\Omega
({\HI})_{\rm outflow} =8.57^{+0.86}_{-1.10}\times 10^{-5}$.  The
former is 6\% of $\Omega_{\hbox{\tiny DLA}}$ and 0.1\% of $\Omega_b$,
and latter is 9\% of $\Omega_{\hbox{\tiny DLA}}$ and 0.2\% of
$\Omega_b$.
%We find a slightly lower neutral gas mass density for
%infalling/accreting material.  
The range of $W$ over which the infall and outflow $\Omega({\HI})$ are
statistically consistent is $W=1.23^{+0.15}_{-0.28}$~{\AA}.

There is no {\it a priori\/} expectation that our approach to
computing $\Omega({\HI})_{\rm halo}$ should yield $\Omega({\HI})_{\rm
infall} \simeq \Omega({\HI})_{\rm outflow}$ for $W \simeq W_*$.  Our
result may indicate that, over a redshift range covering a large
percentage of the age of the universe in the ``post star forming
era'', a cyclic balance persists between inflow and outflow of
galaxies whereby star formation is fueled by accreting gas and then an
equal mass of gas is ejected back into the halos.  This is quite
suggestive of a halo gas recycling model
\citep[e.g.,][]{oppenheimer08}.

% at $z>2.2$
%[for {\HI} selected gas, where $\Omega({\HI})_{<20.3}$ is for
%$N({\HI})<10^{20.3}$~{\cmsq}] \citep{noterdaeme09,peroux05}

\citet{noterdaeme09} extrapolated their {\HI} column density
distribution below $\log N({\HI}) = 20.3$ and found that LLSs
contribute $\simeq$$13$\% of the total $\Omega(\HI)$ at $z>2.2$
\citep[also see][]{peroux05}.  If we assume $\Omega({\HI})_{\rm tot} =
\Omega({\HI})_{\rm halo} + \Omega({\HI})_{\hbox{\tiny DLA}}$, we
obtain $\Omega({\HI})_{\rm halo} \simeq 13$\% of $\Omega({\HI})_{\rm
tot}$ for $0.4 \leq z \leq 1.4$; since $\Omega({\HI})_{\hbox{\tiny
DLA}}$ is constant with redshift, this suggests that
$\Omega({\HI})_{\rm halo}$ has remained constant with redshift and
implies that the {\HI} mass cycling through halos via infall/outflow
has also remained constant.

Thus, given the cosmic evolution of the global SFR (especially below
$z \simeq 2$), and presuming galactic infall/outflow is strongly
coupled to star formation, the global SFR must be governed by the {\it
rate\/} at which {\HI} gas cycles through halos (i.e., the SFR and
{\HI} halo cycling rate must evolve in parallel).  The observation
that the mean ionization of {\MgII} absorbers has decreased with
decreasing redshift from $z\sim 2$ \citep{bergeron94} is consistent
with this scenario.  At higher redshift, the more highly ionized {\HI}
halo gas selected by {\MgII} absorption constitutes a smaller fraction
of the total gas {\it associated with\/} {\HI}; at lower redshift, the
more neutral {\HI} gas constitutes a larger fraction of the total gas
mass.  Thus, the total gas mass associated with {\HI} that is cycling
through halos is higher at high redshift and lower at low redshift for
a fixed {\HI} mass.  Based upon simulations, it is predicted that the
total halo gas mass increases with decreasing redshift to the present
epoch \citep[e.g.,][]{dave99}, however, this is not inconsistent with
our proposed scenario because this growth is in the ``hot'' phase that
is neither detected via {\MgII} absorption nor a reservoir for
star formation.

\citet{ribaudo11} presented possible observational evidence of 
cold accretion in a $[\hbox{Mg/H}] =-1.7$ LLS at $z=0.27$ near a
$Z\simeq Z_{\odot}$ sub-$L_*$ galaxy.  Cosmological simulations
predict that cold accretion is truncated at low redshifts
\citep[e.g.,][]{fumagalli11,stewart11b} such that the cross section of
this gas is a tiny fraction of the observed {\MgII} cross section
\citep{kacprzak08,chen10a}.  If cold, metal-poor filaments comprise a
component of the infalling material, our findings imply they
constitute no more than $\sim 7$\% of $\Omega ({\HI})_{\rm tot}$ at
$\left<z\right>=1.0$.

Our calculation of $\Omega ({\HI})_{\rm halo}$ relies heavily on the
statistical $N({\HI})$--$W$ relation of \citet{menard09b}, which we
extrapolated to $W=0$ and $W=4.5$~{\AA}.  For our calculations of
$\Omega ({\HI})_{\rm infall}$ and $\Omega ({\HI})_{\rm outflow}$, we
assumed the break in the $n(W)$ Schechter function at $W_*\simeq
1$~{\AA} is due to infall/accretion for $W<W_*$ and outflowing/winds
for $W>W_*$.  A bifurcation at $\simeq 1$~{\AA} in the equivalent
width distribution separating the two physical processes of infall and
outflow is an intriguing result that is not well understood within the
framework of current models.  In reality, some fraction of the
$W<1$~{\AA} absorbers could arise in winds and some fraction of the
$W>1$~{\AA} absorbers could arise in infalling/accreting material.
Nonetheless, the observational data suggest that the {\it majority\/}
of $W>1$~{\AA} are of wind origin, and visa versa.  Since there are no
data to constrain the possible fractional contribution of infall or
outflow that may reside on either side of $W_*$, we make no attempt to
quantitatively estimate it.

Further observations are required to ascertain the veracity of this
simple scenario.  For example, there is mounting evidence that galaxy
orientation plays some role in determining $W$ and its origin
\citep{kacprzak11b,bordoloi11}. Although, the evidence provided by
\citet{kacprzak11b} supports the idea that $W\lesssim 1$~{\AA} systems
trace accreting halo gas.

\section{Conclusion}

We have shown that the {\MgII} equivalent width distribution, $n(W)$,
at $\left<z\right>=1.0$, is well described by a Schechter function.
We combined our $n(W)$ with the $N({\HI})$--$W$ relation of
\citet{menard09b} to compute $\Omega ({\HI})$ residing in galactic
halos, as traced by {\MgII} absorption (excluding DLAs). We found that
13\% of $\Omega ({\HI})_{\rm tot}$ resides in galaxy halos and deduced
that the infall and outflowing components comprise roughly equal {\HI}
mass contributions.  The balance between the two may suggest that
outflows are sustained by accretion and that cold accretion by
filaments comprises less than $\sim 7$\% of $\Omega({\HI})_{\rm tot}$.
Comparing to high redshift results, it appears that $\Omega
({\HI})_{\rm halo}$ has not strongly evolved over cosmic time.  We
argued that this implies that evolution in the cosmic SFR must depend
primarily on the rate at which cool {\HI} gas cycles through halos,
even through the total {\HI} mass cycling through halos remains fairly
constant.

%%%%%%%%%%%%%%%%%%%%%%%%%%%%%%%%%%%%%%%%
\acknowledgments 

We thank Dan Nestor for providing data in electronic form and Michael
T. Murphy for carefully reading this paper. We thank the anonymous
referee for providing insightful comments that improved this
Letter. CWC gratefully acknowledges support by Swinburne Faculty
Research Grant, GGK, and Michael T. Murphy during his stay at
Swinburne University of Technology.

%% Included in this acknowledgments section are examples of the
%% AASTeX hypertext markup commands. Use \url without the optional [HREF]
%% argument when you want to print the url directly in the text. Otherwise,
%% use either \url or \anchor, with the HREF as the first argument and the
%% text to be printed in the second.

%doing the math in section~\ref{bozomath}.
%More information on the AASTeX macros package is available \\ at
%\url{http://www.aas.org/publications/aastex}.
%For technical support, please write to
%\email{aastex-help@aas.org}.


%% in parentheses, after the keyword, but they will not be verified.

%{\it Facilities:} \facility{HST (WFPC--2)}, \facility{Keck II (ESI)},
%\facility{Keck I (HIRES)}, \facility{VLT (UVES)}.


\begin{thebibliography}{}

\bibitem[Bergeron {\etal}(1994)]{bergeron94} Bergeron, J., Petitjean, P.,
Sargent, W.~L.~W., {\etal} 1994, \apj, 436, 33

%\bibitem[Bond {\etal}(2001a)]{bond01a} % superbubbles
%Bond, N. A., Churchill, C. W., Charlton, J. C., \& Vogt, S. S. 2001a, 
%ApJ, 557, 761

%\bibitem[Bond {\etal}(2001)]{bond01b} % high-z winds 
%Bond, N. A., Churchill, C. W., Charlton, J. C., \& Vogt, S. S. 2001,
%ApJ, 562, 641

\bibitem[Bordoloi {\etal}(2011)]{bordoloi11} 
Bordoloi, R., Lilly, S.~J., Knobel, C., {\etal} 2011, arXiv:1106.0616

\bibitem[Bouch\'{e} {\etal}(2006)]{bouche06}
Bouch\'{e}, N., Murphy, M. T., P\'{e}roux, C., Csabai, I. \& Wild. V. 2006 MNRAS, 371, 495
\bibitem[Bouwens et al.(2011)]{bouwens11} 
Bouwens, R.~J., Illingworth, G.~D., Labbe, I., et al.\ 2011, \nat,
469, 504

\bibitem[Chelouche \& Bowen(2010)]{chelouche10}
Chelouche, D., \& Bowen, D.~V.\ 2010, \apj, 722, 1821 

\bibitem[Chen {\etal}(2010b)]{chen10b} 
Chen, H.-W., Wild, V., Tinker, J.~L., Gauthier, J.-R., Helsby,
J.~E., Shectman, S.~A., \& Thompson, I.~B.\ 2010b, \apjl, 724, L176

\bibitem[Chen {\etal}(2010a)]{chen10a} Chen, H.-W., Helsby, 
J.~E., Gauthier, J.-R., Shectman, S.~A., Thompson, I.~B., 
\& Tinker, J.~L.\ 2010a, \apj, 714, 1521 

%\bibitem[Chen \& Tinker(2008)]{chen08} 
%Chen, H.-W., \& Tinker, J.~L.\ 2008, ApJ, 687, 745 

\bibitem[Churchill, Kacprzak, \& Steidel(2005)]{cwc-china}
Churchill, C. W., Kacprzak, G. G., \& Steidel, C. C. 2005, 
IAU Proc., 199, 24

\bibitem[Churchill {\etal}(2000)]{archiveI}
Churchill, C. W., Mellon, R. R., Charlton, J. C., Jannuzi, B. T.,
Kirhakos, S., Steidel, C. C., \& Schneider, D. P. 2000, ApJS, 130, 91

\bibitem[Churchill {\etal}(1999)]{weakI}
Churchill, C. W., Rigby, J. R., Charlton, J. C., \& Vogt, S. S. 1999, ApJS, 120, 51

%\bibitem[Churchill \& Vogt(2001)]{cv01}
%Churchill, C. W., \& Vogt, S. S. 2001, AJ, 122, 679 

\bibitem[Dav{\'e} {\etal}(1999)]{dave99} Dav{\'e}, R., Hernquist, L.,
Katz, N., \& Weinberg, D.~H.\ 1999, \apj, 511, 521

\bibitem[Dav{\'e} {etal}(2010)]{dave10} Dav{\'e}, R., Oppenheimer,
B.~D., Katz, N., Kollmeier, J.~A., \& Weinberg, D.~H.\ 2010, \mnras,
408, 2051

\bibitem[Fumagalli et al.(2011)]{fumagalli11} 
Fumagalli, M., Prochaska, J.~X., Kasen, D., et al.\ 2011, \mnras, 1589

\bibitem[Jarosik {\etal}(2011)]{wmap-7yr} Jarosik, N., {\etal} 2011,
ApJS, 192, 14

\bibitem[Kacprzak {\etal}(2011a)]{kacprzak11a} 
 Kacprzak, G.~G., Churchill, C.~W., Barton, E.~J., \& Cooke, J.\ 2011a,
\apj, 733, 105

\bibitem[Kacprzak {\etal}(2011b)]{kacprzak11b} 
Kacprzak, G.~G., Churchill, C.~W., Evans, J.~L., Murphy, M.~T., \&
Steidel, C.~C.\ 2011b, \mnras, 1239

\bibitem[Kacprzak {\etal}(2008)]{kacprzak08} 
Kacprzak, G.~G., Churchill, C.~W., Steidel, C.~C., \& Murphy, M.~T.\
2008, AJ, 135, 922

\bibitem[Kacprzak {\etal}(2010a)]{kacprzak10a} 
Kacprzak, G.~G., Churchill, C.~W., Ceverino, D., Steidel, C.~C.,
Klypin, A., \& Murphy, M.~T.\ 2010a, ApJ, 711, 533

\bibitem[Lah {\etal}(2008)]{lah08} Lah, P., Chengalur, J.~N., Briggs,
F.~H., {\etal} 2008, IAU Symp., 244, 366

%\bibitem[Madau {\etal}(1998)]{madau98} Madau, P., Pozzetti, L., 
%\& Dickinson, M.\ 1998, \apj, 498, 106 

\bibitem[Martin \& Bouch{\'e}(2009)]{martin09} 
Martin, C.~L., \& Bouch{\'e}, N.\ 2009, \apj, 703, 1394 

\bibitem[Martin {\etal}(2010)]{martin10} 
Martin, A.~M., Papastergis, E., Giovanelli, R., {\etal} 2010, \apj,
723, 1359

\bibitem[Meiring {\etal}(2011)]{meiring11} 
Meiring, J.~D., Tripp, T.~M., Prochaska, J.~X., {\etal} 2011, \apj,
732, 35

\bibitem[M{\'e}nard \& Chelouche(2009)]{menard09b}
M{\'e}nard, B., \& Chelouche, D.\ 2009, \mnras, 393, 808

%\bibitem[M{\'e}nard {\etal}(2009)]{menard09a} 
%M{\'e}nard, B., Wild, V., Nestor, D., Quider, A., \& Zibetti, S.\
%2009, arXiv:0912.3263

\bibitem[Narayanan {\etal}(2007)]{narayanan07} 
Narayanan, A., Misawa, T., Charlton, J.~C., \& Kim, T.-S.\ 2007, ApJ,
660, 1093

\bibitem[Nestor {\etal}(2011)]{nestor11} 
Nestor, D.~B., Johnson, B.~D., Wild, V., {\etal} 2011, \mnras, 412,
1559

\bibitem[Nestor {\etal}(2005)]{nestor05} 
Nestor, D.~B., Turnshek, D.~A., \& Rao, S.~M.\ 2005, ApJ, 628, 637 

\bibitem[Noterdaeme {\etal}(2009)]{noterdaeme09} 
Noterdaeme, P., Petitjean, P., Ledoux, C., \& Srianand, R.\ 2009,
\aap, 505, 1087

\bibitem[Noterdaeme {\etal}(2010)]{noterdaeme10} 
Noterdaeme, P., Srianand, R., \& Mohan, V.\ 2010, \mnras, 403, 906

\bibitem[Oppenheimer \& Dav{\'e}(2008)]{oppenheimer08} 
Oppenheimer, B.~D., \& Dav{\'e}, R.\ 2008, \mnras, 387, 577 

\bibitem[Penton {\etal}(2004)]{penton04} Penton, S.~V., Stocke, 
J.~T., \& Shull, J.~M.\ 2004, \apjs, 152, 29 

\bibitem[P{\'e}roux {\etal}(2005)]{peroux05} 
P{\'e}roux, C., Dessauges-Zavadsky, M., D'Odorico, S., Sun Kim, T., \&
McMahon, R.~G.\ 2005, \mnras, 363, 479

\bibitem[Prochaska(1999)]{prochaska99} Prochaska, J.~X.\ 1999, \apjl,
511, L71

\bibitem[Prochaska \& Herbert-Fort(2004)]{prochaska04}
Prochaska, J.~X., \& Herbert-Fort, S.\ 2004, \pasp, 116, 622

\bibitem[Rao {\etal}(2006)]{rao06} 
Rao, S.~M., Turnshek, D.~A., \& Nestor, D.~B.\ 2006, \apj, 636, 610

%\bibitem[Rao \& Turnshek(2000)]{rao00} 
%Rao, S.~M., \& Turnshek, D.~A.\ 2000, ApJs, 130, 1 

\bibitem[Ribaudo {\etal}(2011)]{ribaudo11} Ribaudo, J., Lehner, 
N., Howk, J.~C., {\etal} 2011, arXiv:1105.5381 

%\bibitem[Rigby, Charlton, \& Churchill(2002)]{weakII}
%Rigby, J. R., Charlton, J. C., \& Churchill, C. W. 2002, ApJ, 565, 743

%\bibitem[Rubin {\etal}(2010a)]{rubin10a} 
%Rubin, K.~H.~R., Prochaska, J.~X., Koo, D.~C., Phillips, A.~C., \&
%Weiner, B.~J.\ 2010, \apj, 712, 574

\bibitem[Rubin {\etal}(2010b)]{rubin10b}
 Rubin, K.~H.~R., Weiner, B.~J., Koo, D.~C., {\etal} 2010, \apj, 719,
1503

\bibitem[Schechter(1976)]{schechter76} 
Schechter, P.\ 1976, \apj, 203, 297

\bibitem[Songaila \& Cowie(2010)]{songaila10} 
Songaila, A., \& Cowie, L.~L.\ 2010, \apj, 721, 1448

\bibitem[Steidel {\etal}(2002)]{steidel02} 
Steidel, C. C., Kollmeier, J. A., Shapely, A. E., Churchill, C. W.,
Dickinson, M., \& Pettini, M. 2002, ApJ, 570, 526

\bibitem[Steidel \& Sargent(1992)]{steidel92} 
Steidel, C.~C., \& Sargent, W.~L.~W.\ 1992, ApJS, 80, 1 

\bibitem[Stengler-Larrea {\etal}(1995)]{stengler-kp95} 
Stengler-Larrea, E.~A., Boksenberg, A., Steidel, C.~C., {\etal} 1995, \apj, 
444, 64 

\bibitem[Stewart {\etal}(2011a)]{stewart11a} 
Stewart, K.~R., Kaufmann, T., Bullock, J.~S., {\etal} 2011a, \apj, 738,
39

\bibitem[Stewart {\etal}(2011b)]{stewart11b} 
Stewart, K.~R., Kaufmann, T., Bullock, J.~S., {\etal} 2011b, \apjl, 735, L1

\bibitem[Tremonti {\etal}(2007)]{tremonti07} 
Tremonti, C.~A., Moustakas, J., \& Diamond-Stanic, A.~M.\ 2007, ApJl,
663, L77

\bibitem[Weiner {\etal}(2009)]{weiner09} 
Weiner, B.~J., {\etal} 2009, ApJ, 692, 187

\bibitem[Wolfe {\etal}(2005)]{wolfe+05} 
Wolfe, A.~M., Gawiser, E., \& Prochaska, J.~X.\ 2005, ARA\&A, 43, 861 

\bibitem[Zibetti {\etal}(2007)]{zibetti07}
Zibetti, S., M{\'e}nard, B., Nestor, D.~B., Quider, A.~M., Rao, S.~M., \& Turnshek, 
D.~A.\ 2007, ApJ, 658, 161 

\bibitem[Zwaan {\etal}(2005a)]{zwaan05a} 
Zwaan, M.~A., Meyer, M.~J., Staveley-Smith, L., \& Webster, R.~L.\
2005a, \mnras, 359, L30

\bibitem[Zwaan et al.(2005b)]{zwaan05b} 
Zwaan, M.~A., van der Hulst, J.~M., Briggs, F.~H., Verheijen,
M.~A.~W., \& Ryan-Weber, E.~V.\ 2005b, \mnras, 364, 1467

\end{thebibliography}
\end{document}